# Homodyne detection for quantum key distribution: an alternative to photon counting in BB84 protocol

M.B.Costa e Silva[a], Q. Xu[a], S. Agnolini[a], P. Gallion[a], F.J. Mendieta[b]

[a]Ecole Nationale Supérieure de Télécommunications (GET/Télécom Paris and CNRS LTCI), 46 rue Barrault, 75013 Paris, France.
[b]CICESE, km. 107 Carr. Tijuana, Ensenada, Baja California 22800, México.

## ABSTRACT

This paper presents the principles and experimental results of an optical fiber QKD system operating at 1550 nm, and using the BB84 protocol with QPSK signals. Our experimental setup consists of a time-multiplexed super-homodyne configuration using P.I.N detectors in a differential scheme as an alternative to avalanche photon counting. Transmission over 11km of optical fiber has been done using this detection scheme and major relevant characteristics such as noise, quantum efficiency and bit error rate (BER) are reported.

## 1. INTRODUCTION

The first experiments of quantum key distribution (QKD) using fainted sources (attenuated light pulses), were carried out at visible or near infrared wavelengths, based mainly on the polarization of the photons. More recently, experiments at telecommunications wavelengths have been reported, introducing the coding other than polarization, such as the optical phase [1], and detection by photon counting with cooled avalanche photodiodes (APD) [2]. However these receivers present severe limitations in the present 1550 nm systems, due to their inherent low quantum efficiency and their high dark count rate, which requires an operation in the gated mode, resulting in a very low key generation rate even at low operation temperature [3].

To obtain substantially higher key rates, homodyne detection constitutes an interesting alternative to photon counting, because when used with a local oscillator of suitable power for operation near the quantum noise limit, it provides the mixing gain to overcome the thermal noise, while employing conventional P.I.N. photodiodes operating at room temperature, that present a much higher quantum efficiency and response speed as compared to APD, as well as lower cost and simpler supply requirements [4]. Homodyne detection has already been investigated to provide accurate quadrature measurements in QKD using continuous random variables [5].

Furthermore, the coherent detection technique allows a diversity of modulation formats of the optical carrier, which is important when multiple states of phase are required, such as MPSK for the quantum bit (Q-bit) modulation, like in the BB84 protocol. Thus the objective of our work is to use this type of reception in a super-homodyne configuration for this protocol.

When homodyne reception is used for the detection of Q-bits, a balanced configuration must be carried out to reach the necessary quantum measuring accuracies; therefore one must extract, at Bob's end, a reference of the optical carrier, to generate the local field that provides an acceptable mixing gain. Furthermore the receiver must be designed to accommodate with the random phase fluctuations in the channel due to the optical source at Alice's end (in general fast) and to the thermo-mechanical states of fibers and other in-line components (in general slow, but introducing depolarizing effects as well).

Post-detection, filtering, threshold and symbol synchronization stages must also be properly designed as in homodyning the logical decision process is carried out after balanced photodetection, as opposed to photon counting that inherently performs built-in decision. The former leads to classical bit error rate (BER) while only quantum bit error rate (QBER) is considered in the latter.

In this work we encode the Q-bits in two orthogonal bases with two antipodal symbols per base, leading to a QPSK modulation format; and we present two configurations for its measurement: first a self homodyne system that transmits a strong unmodulated carrier on a separate fiber as the demodulation reference; and also a single fiber system that introduces high power pulses of the carrier reference, time multiplexed with the faint pulses, followed by a delayed interferometric homodyne reception. This differential configuration relaxes the polarization and phase fluctuations conditions in the common channel, since an absolute stability is not required but only a relative stability during a few symbol periods.

## 2. QKD WITH QPSK FORMAT

### 2.1 BB84 protocol

In cryptographic communication it is mandatory to send the encrypted message and in a secure way the key to decrypt it. Quantum key distribution (QKD) permits the exchange of the key in a "quantum" channel, between Alice and Bob. This key has to be protected from Eve that tries to intercept the message. The protocol is a group of strict rules that is indispensable for a QKD system to be implemented as an unconditionally secure communication.

The protocol proposed by Bennett and Brassard in 1984 (BB84) [6] uses some basic quantum concepts to operate. From two orthogonal bases chosen randomly by Alice, four quantum eigen states can be generated separately (the symbols 0 and 1 on two different bases). Hence for the communication, Alice randomly chooses her symbol with a random base before sending it to Bob; then Bob also chooses his base to read the received bit. Bob and Alice then talk in a classical channel to compare the chosen bases after a sequence of key bits have been received; a raw key is thus generated.

When there is base coincidence, i.e., Bob and Alice choose the same base, the bit is correctly detected and when there is anti-coincidence, the measurement is random. The sequence obtained when there are base coincidences is kept, and then some of these bits are chosen to perform the eavesdropping test, i.e. privacy amplification. Alice and Bob compare these symbols of raw keys to obtain a sifted key, which is then used for the encryption of the message. "One-Time Pad" (Vernam Code) is usually employed in this kind of system as to avoid any eventual information leakage.

To guarantee the security in this protocol it is necessary that the communication be done at quantum levels, ideally 1 photon per transmitted bit. This can provide the unconditional security, since if Eve tries to read this unique photon, Bob will be aware of it due to quantum mechanics principles, otherwise if Eve reads it and the retransmits it, the error is easily detected by the reconciliation procedure between Alice and Bob and the bit is discarded. [6]

### 2.2 Generating QPSK for QKD

The BB84 protocol requires Alice's choice from two bases, and each base has two symbols. This permits four different parameter values. In an optical fiber scheme operating with phase modulation, the symbols must have antipodal phase states in two conjugated bases, and the BB84 requirements can be met by positioning each of these four values as one of four points in a QPSK constellation.

Hence Alice generates 4 different phase states to perform this task, our QKD system utilizes a two-electrode Mach-Zehnder electro-optical modulator (EOM), permitting the independent control in each electrode arm. With this kind of modulator the optical field undergoes variation both in amplitude and phase, depending on the signals introduced to each arm following the equation:

$$E_1(t) = E_0(t) \cdot \cos\left(\frac{\phi_1 - \phi_2}{2}\right) \exp\left(j\frac{\phi_1 + \phi_2}{2}\right) \qquad (1)$$

where $\phi_1$ and $\phi_2$ are the phase shifts induced by the modulation tension applied in electrode 1 and 2 respectively, as shown in Fig. 1(a).

In order to generate the QPSK signals and keep the intensity constant, i.e. $\phi_1 - \phi_2 = \pm \pi/2$, an arrangement to apply this Alice's modulation has been done as shown in Fig. 1(a). Once the base and symbol choices are made these two signals are added each in one arm of Alice's modulator. The optical field of Alice's modulation is:

$$E_1(t) = E_{A0}(t) \cdot \exp\left(j\frac{\phi_1 + \phi_2}{2}\right) \qquad (2)$$

Bob also chooses his base, operating with the same kind of modulator as Alice (EOM-A for Alice and EOM-B for Bob), using only one electrode input, adds to the field a new phase variation permitting to extract the base choices and symbol information as established in the BB84 protocol. A table of Alice's choices of bases and symbols and Bob's choices of bases, as well as the key coincidence/anti-coincidence are shown in Fig. 1 (b) [7].

$$E_1(t) = E_{B0}(t) \cdot \cos(\phi_B) \cdot \exp(j(\phi_A + \phi_B)) \qquad (3)$$

where $\phi_A = \frac{\phi_1 + \phi_2}{2}$, $\phi_B = \frac{\phi_3 + \phi_4}{2}$, but $\phi_4$ is not used in the scheme.

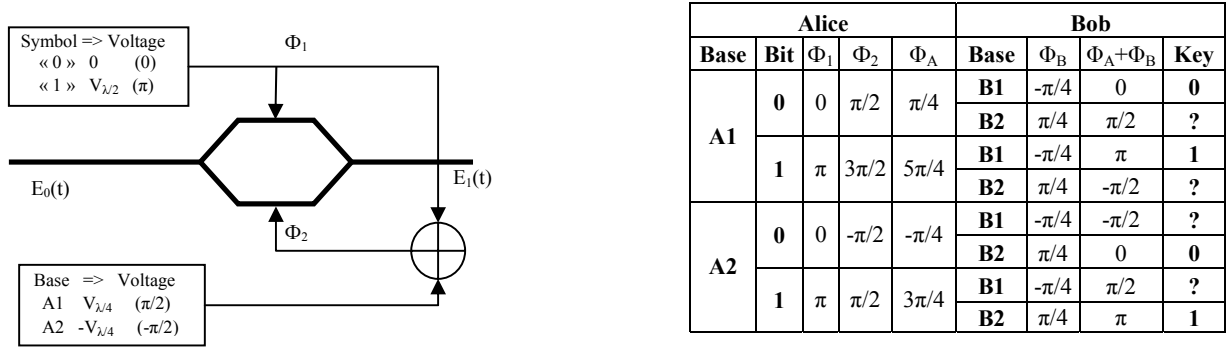

Fig. 1 (a) Alice's encoding diagram (b) table for QPSK BB84 protocol

## 3. EXPERIMENTAL SET-UP

### 3.1 Self-homodyne receiver

Our first experiment to validate this QPSK QKD system consists of a standard telecommunications optical-fiber self-homodyne system with a strong carrier "reference" transmitted in a separate optical line. The modulated signal arm was constructed to have EOM-A followed by an optical attenuator; Bob introduces his base choices in the reference arm at the reception, the setup is shown in Fig. 2. Alice and Bob apply their respective modulation signals, obtaining the same results and keys as in Fig. 1(b). The curve in Fig. 3(a) shows the detected signal when a square electrical signal is used for the modulation, the coincidence of bases between Alice and Bob are shown in the waveform as positive or negative levels, while the anti-coincidence are discarded (level 0); furthermore from the histogram, Fig. 3(b), we can observe the manifestation of the base coincidences (outer peaks) and anti-coincidences (inner peaks).

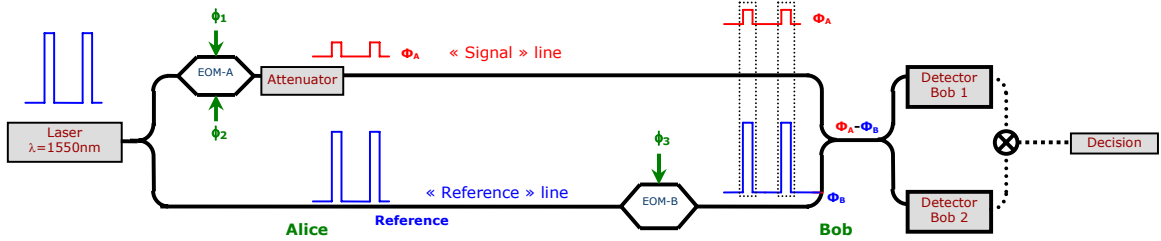

Fig.2 QKD QPSK Self-Homodyne Setup

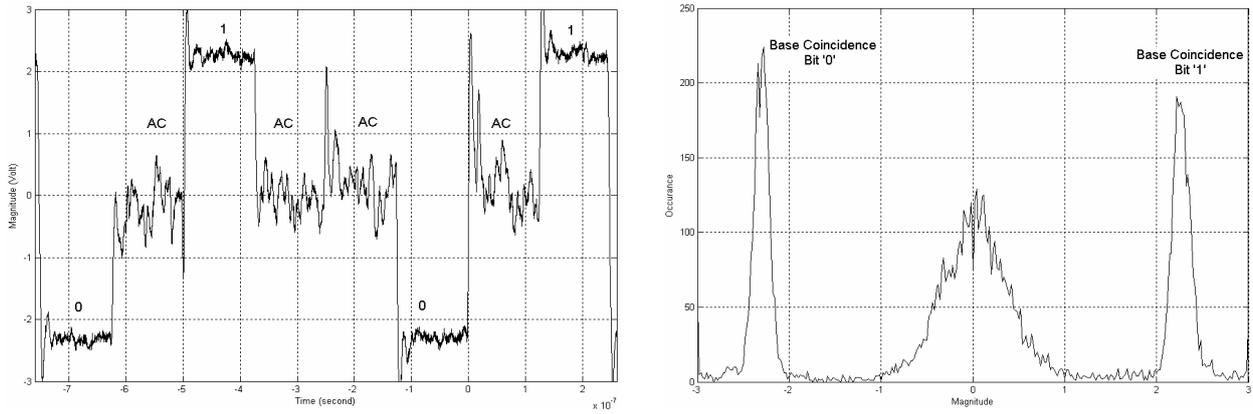

Fig. 3 (a) Detected Output where AC= anti-coincidence and (b) Histogram

Strictly the protocol requires, for unconditional security, that the transmission is performed with 1 photon per bit, thus requiring a (non-existing yet) single photon source. In practice we use an attenuated DFB source, which follows the Poissonian statistics, and a security level can be reached when an approximation of 0.1 photon/bit is used [5], at the cost of a quantity of empty pulses and of multi-photon pulses. Consequently we place the attenuator after Alice's modulation to prepare the optical signal to a quantum level in order to assure the required security; in addition modulating the laser with narrow pulses can help provide a much lower number of photons per bit. Fig. 4 is an example of the same setup but with very weak optical level: the modulation was introduced as before. The waveform that can be seen by the oscilloscope, Fig. 4(a), shows the positive or negative pulses when there are base coincidences and zero levels when there are anti-coincidences. In addition from a histogram of received signals, Fig. 4(b), we can obtain the bit occurrence; and this time the base anti-coincidence bits are immersed in the noise. This result was obtained when the optical power was attenuated from $4.83\times10^7$ photon/bit (as shown in Fig. 3) to $3.8\times10^4$ photon/bit.

The first self-homodyne configuration requires the transmission in two fibers, which suffer from unequal phase and polarization fluctuations as the propagation distance increases. In addition, the "signal" and "reference" lines must be in absolute phase-alignment, which is technically very difficult to implement in a practical application.

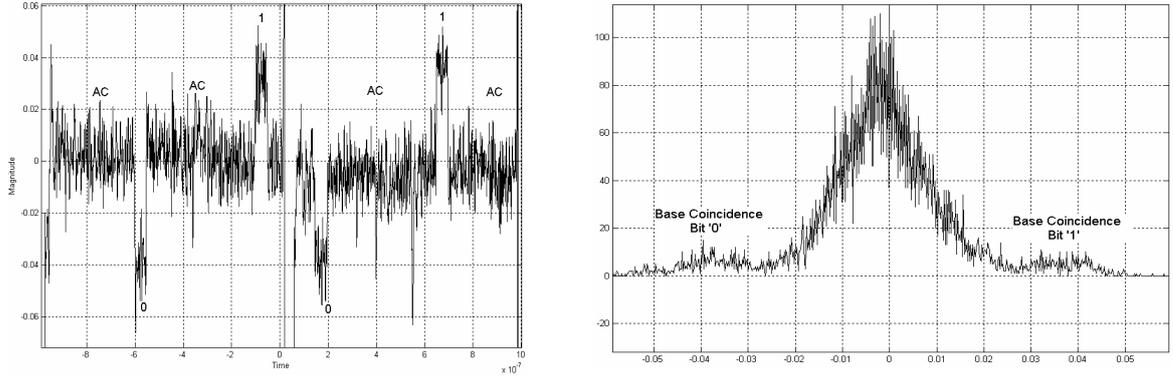

Fig. 4 (a) Detected Output where AC = anti-coincidence (b) Histogram for Pr = $3.8 \times 10^4$ photon/bit

### 3.2 The time-multiplexed Super-Homodyne configuration

Our second homodyne configuration consists of sending the weak QPSK-modulated pulses time-multiplexed with the unmodulated strong pulses that constitute a carrier phase reference in a same fiber [8]. Fig. 5 is a diagram of our experimental setup: the coherent optical pulses are fed into Alice's unbalanced interferometer, then Alice's fainted pulses are produced in her longer arm with EOM-A as mentioned above, while strong unmodulated pulses pass through the shorter arm, with accurate polarization control. The pulses are recombined after the interferometer and pass through 11 km fiber arriving at Bob's. The Alice's detector as in Fig. 5 is used for monitoring the polarization control.

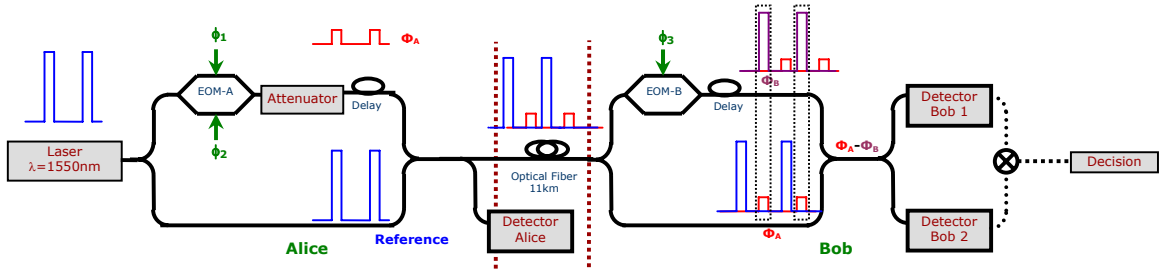

Fig. 5 QKD DQPSK Setup: Coherent Super-Homodyne Balanced Detection

At the receiver, Bob's measurements are performed by applying his 2-state phase modulation to the strong pulses in a similar delayed interferometric configuration so that the weak key pulses beat with the high power reference pulses in order to achieve the mixing gain; then a balanced configuration is used for photodetection.

The extinction ratio is a determinant element in the time-multiplexing setup. The pulses in the "signal" line are very weak in QKD application, generally 0.1—1 photon/symbol; therefore the finite extinction ratio of the "reference" pulses must be sufficiently high.

Actually, the signal and the reference are set to mutually orthogonal polarizations when they are combined (at Alice) so as to avoid interference; and rendered to the same polarizations when beating with each other (at Bob) in order to maximize the detected signal. Phase stability in the interferometer is another important factor which is dependant of the temperature variation and the internal and external mechanical vibrations. The unbalanced interferometer is several meters long in our experimental setup.

## 4. MEASUREMENTS

As Alice transmits single photon pulses and encodes the information in their phases, the detection of signal photons is essential for a practical QKD system. Because any background light in the fiber can increase the system noise level and limit its performance, high-sensitivity receivers are thus required for detecting the very weak signals. Photon counting and homodyne detection are two available methods for detecting weak light. We will compare their characteristics and performances in this section. One technical limitation of the photon counting method is that at present there exists no perfect single photon source and no efficient photon counter either for telecom wavelength where optical loss is minimized in an optical fiber.

### 4.1 Photon counting receivers

As a preliminary experiment, we present a receiver based on photon counters as a reference for comparison with our homodyne detection system.

The usual way to improve signal photon sensitivity of an APD is to operate in a Geiger Gate mode, when it is biased above the breakdown. A single photon can then trigger an avalanche containing millions of electrons and create a macroscopic current pulse. After a detection event, avalanche is quenched and gated to decrease the dark count and prepare the APD for the next multiplication, thus making the detection of single photon events possible.

In this experimental setup, we use a single-photon detector module (SPDM from ID Quantique) which offers a dark count below 100 counts per second and sub-nanosecond timing resolution [9].

Fig.6 is an experimental setup for the photon counting system. The pulsed laser is attenuated to a quantum level before entering the system. Since EOM-A and EOM-B have the same insertion loss, we apply Alice's and Bob's phase modulations both in the longer arm of the two unbalanced interferometers so that the pulses of $\Phi_A$ and those of $\Phi_B$ have the same intensity. Detector 1 clicks for $\Phi_A-\Phi_B = 0$ while Detector 2 clicks for $\Phi_A-\Phi_B = \pi$. When $\Phi_A-\Phi_B = \pi/2$ or $-\pi/2$, the photon arrives at Detector 1 or Detector 2 in a random way.

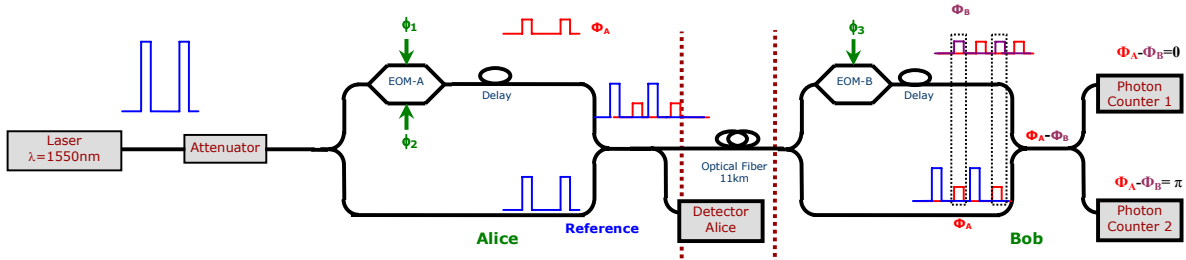

Fig. 6 QKD DQPSK Setup: Photon Counting Detection

In this setup, the repetition frequency is set to 1MHz. The gate width is set to 2.5ns so as to minimize the dark counts which could be randomly triggered by carriers generated in thermal, tunneling or trapping processes taking place in the junction. In fact, the detectors are cooled down to 220K to reduce the occurrence of thermally generated carriers. Another effect to be taken into account is the so-called "afterpulses" which is proportional to the charge crossing the junction in an avalanche before the quenching process, since reducing the operation temperature of the APD increases also the lifetime of the trapped charge. Therefore a compromise must be made between the operation temperature and "deadtime" which inhibits gates for a while. This is also a reason why the maximum frequency of this SPDM is limited to 4MHz. When working at a stable temperature of 220K, the quantum efficiency is lower than 10%, since the dark count rate increases with quantum efficiency if a larger gate width is selected.

When using a fiber interferometer without feedback from the detected signal, a mean false count rate of 30% was obtained in a condition of stable phase for several minutes due to the finite visibility of the interferometer (which leads to a probability of a photon being misdirected to the wrong detector).

The security of practical implementations of the photon counting scheme is limited by: a) the unsatisfactory extinction ratio of the reference pulses; b) the polarization imperfections of the laser pulse which affect the mixing of the weaker key pulse with the strong reference pulse; c) the slight deviation of modulating signals; d) the unavoidable thermo-mechanical variations; and e) the dark counts of the APD detectors.

Intrinsic problems associated with the high gain APD are high amplification noise and dark current noise. The low quantum efficiency and the high probability of a false avalanche due to dark current noise preclude many practical applications. In addition, the single photon source adds to another bottleneck for photon counting detection [9], several research groups are dedicating to produce a practical signal photon source such as heralded single photon source (HSPS) which generates photon pairs. [11]

## 4.2 Balanced Detection

The photocurrent (I=RP, where R is the detector responsivity) resultant of super-homodyne detection [12, 13, 14] is:

$$I_{1,2}(t) = R(P_S + P_{REF}) \pm 2R\sqrt{P_S P_{REF}} \cos(\Phi_A - \Phi_B). \quad (4)$$

When $P_{REF} \gg P_S$, the last term in the equation (4) containing the information transmitted is extracted by the decision circuit. When a balanced detector is used, the homodyne signal is then given by:

$$I(t) = 4R\sqrt{P_S P_{REF}} \cos(\Phi_A - \Phi_B). \quad (5)$$

Denoting the average signal power by $\overline{P_S}$, the average electrical power is increased by a factor of $4P_{REF}/\overline{P_S}$. Although shot noise is also increased, the homodyne detection improves largely the signal-to-noise ratio (SNR), when a suitable reference power is employed. The BER of super-homodyne detection is [12]:

$$BER = \frac{1}{2} erfc\left(\sqrt{2\eta R N_B}\right) \quad (6)$$

where $\eta$ is the quantum efficiency, R is the responsivity, $N_p$ is the number of photons per bit.

Fig.7 shows the theoretical BER for super-homodyne balanced detection assuming that "shot noise" >> "thermal noise" and phase noise is negligible. In fact differential system relaxes substantially the stability requirements in the mean square phase fluctuations $\overline{\Delta\varphi^2(\tau)}$ that increases linearly with time delay τ, as in our delayed configuration τ is of the order of a few bit periods.

Our system uses super-homodyne balanced detection for implementing the Bennett-Brassard BB84 protocol with QPSK coding. We have performed QKD by sending light pulses at 1.55μm through 11km propagation.

As shown in Fig. 5, the light source is a 1.55μm coherent laser source (LMM Digital Electro-Absorption Laser Module from AVANEX) which generates the pulses of 10ns with dynamic extinction ratio > 10dB. Optical pulses are fed into an unbalanced interferometer: in its longer arm we produce Alice's fainted QPSK signal. Strong unmodulated pulses pass through the short arm, as shown in Fig. 5, and are used as reference signal.

Fig. 8(a) shows an example of the combination of weak modulated signal pulses and strong reference signal pulses after Alice's interferometer. The weak modulated signal pulses are delayed as to implement a time-multiplexing configuration.

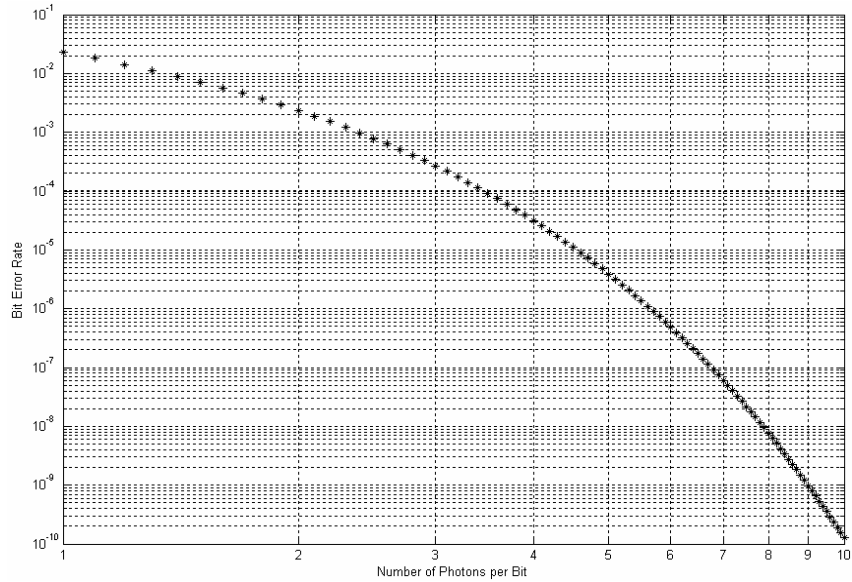

Fig. 7 Theoretical BER of Homodyne Detection

At Bob's end, 2-state phase modulation is applied in a similar delayed configuration so that the weak modulated pulse beats with the strong reference pulse; we use a balanced photoreceiver to perform the detection which has a high sensitivity and a high conversion gain.

The balanced photoreceiver consists of two matched InGaAs photodiodes and a low-noise amplifier that generates an output voltage proportional to $I_2-I_1$, the difference between the photocurrents in the two photodiodes, with transimpedance gain of 40V/ma. Fig. 8(b) shows Bob's detected symbols after 11km fiber propagation using balanced detection.

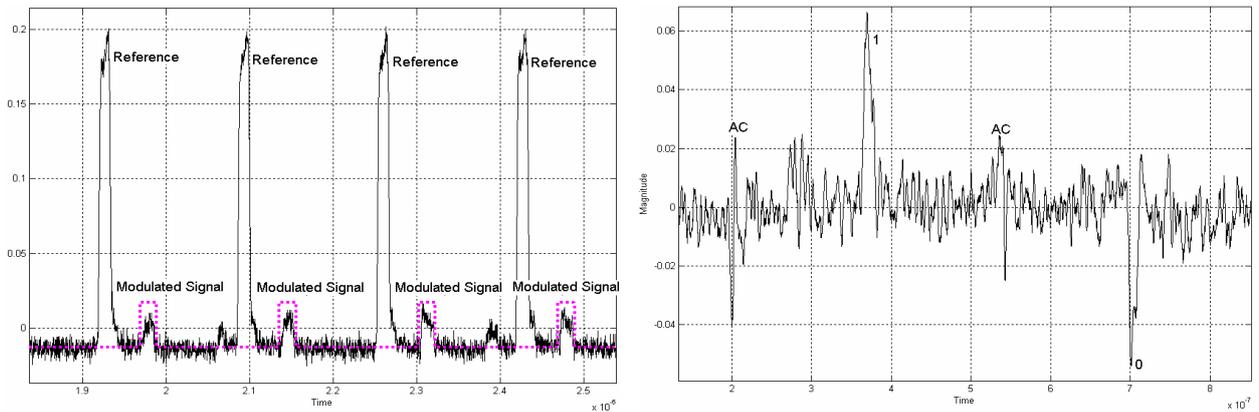

Fig. 8 QPSK Signal (a) Alice's Output (b) Bob's Detected Symbols where AC = anti-coincidence.

The balanced detection has several advantages over photon counting : a) Quantum efficiency of P.I.N detector is near unity which is much higher than photo counters; b) Balanced configuration helps remove common mode noise issued from the electrical circuit of P.I.N detectors, besides it can extract the intensity difference between two received signals and double the received signal amplitude; c) With the proper reference power, shot noise is the only limit since the strong reference pulse makes the thermal noise irrelevant (thermal noise is -174dBm/Hz, whereas in our experimental setup the quantum shot noise is -152dBm/Hz,); d) High speed QKD system is possible with super-homodyne balanced detection since P.I.N diodes do not require quenching process.

Fig. 9(a) demonstrates Bob's detected symbols with 2 photons per symbol, in which the system noise is limited to the shot noise level, as shown in Fig. 9 (b). However, the current fluctuation is still a bottleneck in the balanced detection. As shown in Fig.7 theoretically we can reach a BER less than 0.3% with 2 photons per symbol with super-homodyne detection, as for 1 photon per symbol BER could be around 2%.

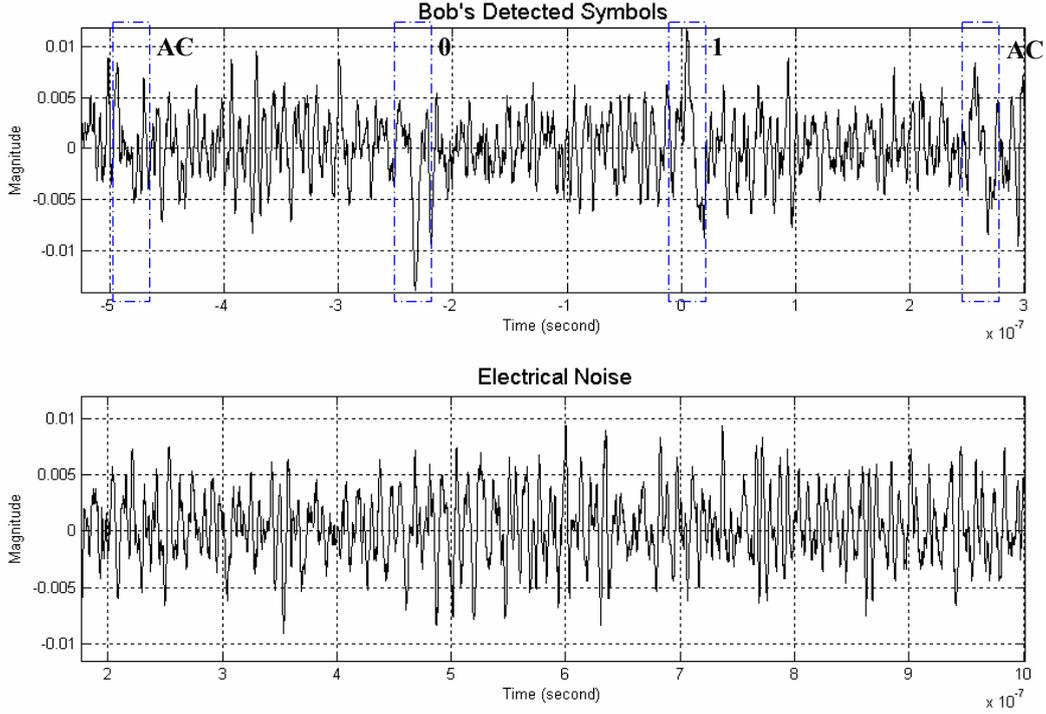

Fig. 9 (a) Bob's detected symbols with 2 photons/bit (b) Bob's detected electrical noise

### 4.3 Photon counting versus homodyne detection

Now we compare the performance of the 2 systems in terms of the BER: For super-homodyne balanced detection, we have [12],

$$SNR = 4\eta_F \eta_B R N_p,  \qquad (7)$$

$$BER = \frac{1}{2} erfc\left(2\eta_F \eta_B R N_p\right) \qquad (8)$$

where $\eta_F$ is the transmission loss (generally 0.2dBm/km, the probability that photons reach detector after propagation through the fiber), where $\eta_B$ is Bob's loss, R is the responsivity, $N_p$ is the number of photons per bit.

For photon counting SPDM [15], we also take into account the detection probability (detector quantum efficiency), the interferometer visibility V, and the dark counts rate $P_{dc}$. The received number of photons/bit is

$$p = \eta_F \eta_B QE_D \times N_P, \tag{9}$$

where $QE_D$ is the quantum efficiency.

$$QBER = \frac{1-V}{2} + \frac{P_{dc}}{2p} + \sum_{i=0}^{n} p_{after}(t_i). \tag{10}$$

The third term of the equation (10) is the afterpulsing probability from the starting avalanche event to the next avalanche event.

Fig. 10 shows the theoretical performance of these two detection configurations over 80km when BER of photon counting configuration is calculated by equation (11), since we consider the erasure of photons as errors. The Dark Count Probability is $10^{-4}$ (100 dark counts/second when the repetition frequency is 1MHz), Bob's loss is 4dB, $QE_D$ is 6%, $N_p$ sent by Alice is 1 photon/bit.

$$BER_{APD} = \frac{QBER}{QE_D} \tag{11}$$

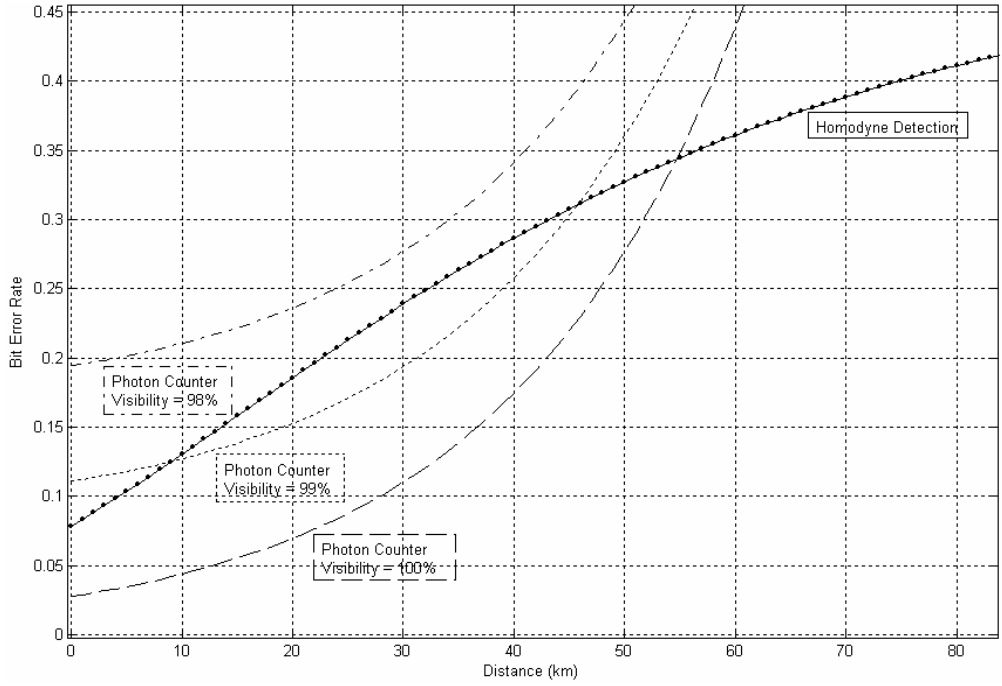

Fig. 10 The performance of photon counting and homodyne detection

The homodyne detection constitutes as a good alternative for QKD system: a) it doesn't exhibit the dark counts which is dominant in photon counting configuration; b) in a practical point of view, its frequency limit is much higher because it is mainly related to the rise time of the InGaAs photodiodes; c) it works at an ambient temperature; d) it allows a variety of key encoding formats (phase and polarization modulation).

## 5. CONCLUSION

An experimental QKD system for the BB84 protocol has been implemented using optical phase modulation for base and symbol encoding, including the transmission of a strong carrier reference that allows a variety of key encoding formats;

we presented a QPSK modulation for the Q-bit with two detection configurations: a self-homodyne with an additional fiber for the strong reference, and a delayed homodyne scheme that uses only one fiber, with a time-multiplexed strong reference pulses. We present the results of measurements for both homodyne configurations using balanced P.I.N. detectors, and, as a comparison, results of the operation with photon counters.

Future work is to be done on the carrier phase synchronization, even in the differential system that requires the quadrature conditions in the interferometer, as well as the polarization handling and the symbol clock synchronization.

## ACKNOWLEDGEMENT

We thank Sylvain Guilley, Jean-Luc Danger for their help in programming the SPDM photon counters; Frédéric Pauget for his technical support in the laboratories. We thank CNPq for financial support of this work for Dr. Marcia Betania Costa e Silva. Dr. Francisco Javier Mendieta thanks ENST for hosting long sabbatical leave from CICESE. Last but not least, we thank AVANEX (Jean-René Buric) for their support of Laser PowerSource 1915, which is used in our QKD system.